\documentclass[twocolumn,pre,showpacs,amsmath,amssymb]{revtex4-1}

\usepackage{amsmath}
\usepackage{dcolumn}
\usepackage{wrapfig}
\usepackage{graphicx}
\usepackage{color}
\usepackage{graphics}
\usepackage{epsfig}
\usepackage{units}
\usepackage{dsfont}
\usepackage{amssymb}
\usepackage{amsthm}
\usepackage{times}

\global\long\def\argmin{\operatornamewithlimits{argmin}}

 \theoremstyle{plain}
  \newtheorem*{thm*}{\protect\theoremname}
  \providecommand{\theoremname}{Theorem}

\begin{document}

\title{Optimal structure and parameter learning of Ising models} 

\author{Andrey Y. Lokhov$^{1,2}$}
\author{Marc Vuffray$^{2}$}
\author{Sidhant Misra$^{3}$}
\author{Michael Chertkov$^{1,2,4}$}
\affiliation{$^{1}$Center for Nonlinear Studies, Los Alamos National Laboratory, Los Alamos, NM 87545, USA,}
\affiliation{$^{2}$Theoretical Division T-4, Los Alamos National Laboratory, Los Alamos, NM 87545, USA,}
\affiliation{$^{3}$Theoretical Division T-5, Los Alamos National Laboratory, Los Alamos, NM 87545, USA,}
\affiliation{$^{4}$Skolkovo Institute of Science and Technology, 143026 Moscow, Russia}

\begin{abstract}
  {\bf One Sentence Summary:} An arbitrary Ising model can be exactly recovered from observations using information-theoretically optimal amount of data.\\
  \linebreak{}
  Reconstruction of structure and parameters of an Ising model from binary samples is a problem of practical importance in a variety of disciplines, ranging from statistical physics and computational biology to image processing and machine learning. The focus of the research community shifted towards developing universal reconstruction algorithms which are both computationally efficient and require the minimal amount of expensive data. We introduce a new method, Interaction Screening, which accurately estimates the model parameters using local optimization problems. The algorithm provably achieves perfect graph structure recovery with an information-theoretically optimal number of samples, notably in the low-temperature regime which is known to be the hardest for learning. The efficacy of Interaction Screening is assessed through extensive numerical tests on synthetic Ising models of various topologies with different types of interactions, as well as on a real data produced by a D-Wave quantum computer. This study shows that the Interaction Screening method is an exact, tractable and optimal technique universally solving the inverse Ising problem.
\end{abstract}

\maketitle

\section*{Introduction}

The Ising model is a renowned model in statistical physics which was originally introduced to study the phase transition phenomenon in ferromagnetic materials \cite{gallavotti2013statistical}. In modern applications, the Ising model is regarded as the most general graphical model describing stationary statistics of binary variables, called spins, that admit a pairwise factorization. The spins are associated with the nodes of a graph and the edges specify pairwise interactions. Given a graph $G=(V,E)$, where $V$ is the set of $N$ nodes and $E$ is the set of edges, the probability measure of an Ising model reads
\begin{equation}
P_{J^{*},H^{*}}(\sigma) = \frac{1}{Z}\exp\left(\sum_{(i,j)\in E} J^{*}_{ij}\sigma_i \sigma_j +\sum_{i\in V}H^{*}_i\sigma_i\right), 
\end{equation}
where $\sigma = \{\sigma_i\}_{i \in V}$ denotes the vector of spin variables $\sigma_{i} \in \{-1,+1\}$, $J^{*} = \{J^{*}_{ij}\}_{(i,j)\in E}$ is the vector of pairwise interactions, $H^{*} = \{H^{*}_i\}_{i\in V}$ is the vector of magnetic fields and $Z$, called the partition function, is a normalization factor that ensures $\sum_\sigma P_{J^{*},H^{*}}(\sigma)=1$. In this representation, the temperature is absorbed in $J^{*}$ and $H^{*}$. Regimes corresponding to small and large interactions and magnetic field intensities are respectively known as high-temperature and low-temperature phases. Models in which couplings or magnetic fields are positive, negative or have mixed signs are traditionally referred to as ferromagnet, anti-ferromagnet and spin glass, respectively. In numerous application fields, such as statistical physics \cite{kunkin1969inverse,chayes1984inverse}, neuroscience \cite{Schneidman2006,Cocco2009}, bio-polymers \cite{Morcos2011a}, gene regulatory networks \cite{Marbach2012}, quantum computing \cite{ronnow2014defining}, image segmentation \cite{panjwani1995markov}, deep learning \cite{lecun2015deep} and sociology \cite{Eagle2009}, the underlying interaction graph and the values of couplings are often unknown \emph{a priori} and have to be reconstructed from the data which takes the form of several observed spin configurations. The learning problem that we consider in this paper, called the inverse Ising problem, is stated as follows: given $M$ statistically independent samples $\{\sigma^{(m)}\}_{m=1,\ldots,M}$ generated by an unknown probability measure $P_{J^{*},H^{*}}(\sigma)$, reconstruct the interaction graph $G$ and the parameters $\{J^{*},H^{*}\}$.

Over the past several decades, a considerable number of techniques have been developed in statistical physics, machine learning and computer science communities in order to carry out this reconstruction task \cite{ackley1985learning,kappen1998boltzmann,Roudi2009,sessak2009small,nguyen2012mean,ricci2012bethe,Cocco2011,sohl2011new,ravikumar2010high,aurell2012inverse,Decelle2014,bresler2008reconstruction,Bresler2015}. A direct maximization of the log-likelihood of the data is generally intractable because it requires a repeated evaluation of the partition function $Z$ for different trial values of the parameters $\{J,H\}$. Computing $Z$ is in general a task of exponential complexity in the number of spins \cite{cooper1990computational}, under exception of some special cases such as tree-structured Ising models \cite{chow1968approximating} and planar Ising models with zero magnetic fields \cite{johnson2015learning}. In spite of this difficulty one may still try computing $Z$ using for instance Monte-Carlo simulations, as done in \cite{ackley1985learning} via the so-called learning for Boltzmann machines. In this method, one estimates all the magnetizations and pairwise correlation functions from samples and then maximizes the log-likelihood using a gradient ascent procedure over all couplings and magnetic fields. The Monte-Carlo nature of the method makes it exponentially expensive in the number of runs required to achieve a pre-defined accuracy. Note, however, that this method is asymptotically exact as the number of samples goes to infinity, thus illustrating that ``sufficient statistics" based approaches that use only estimates of first moments and pair-correlations of spins can achieve exact reconstruction albeit through computations with exponential complexity \cite{montanari2015computational}.

Following the observation that first and second moments are sufficient to reconstruct Ising models, a number of mean-field approximations have been suggested to circumvent the difficulty of an analytical evaluation of magnetizations and pair-correlations functions, see \cite{Roudi2009} for a review. The applicability of these methods is limited: they perform weakly on systems embedded in a low-dimensional space and in the spin glass regime, where fluctuations are important and can not be neglected. Some of the limitations of these na\"{i}ve mean-field methods \cite{kappen1998boltzmann} are addressed in more advanced mean-field methods: the small correlations expansion \cite{sessak2009small} considers corrections to the mean-field in the high-temperature regime; \cite{nguyen2012mean} exploits clustering of samples in the configuration space according to their mutual overlaps; and the Bethe approximation \cite{ricci2012bethe} is based on the tree-like approximation of the interaction graph. Nevertheless the applicability of these approximate techniques remains limited to Ising models pertaining to specific classes.

Although sufficient statistics consisting of the first and second moments of the data carry all the information needed for estimating the couplings, the computations required to extract this information are expensive and prohibitive for large systems \cite{montanari2015computational}. This leaves the use of higher order moments of the spin statistics as the only way to improve computational complexity. Several heuristic algorithms that use higher order statistics have been proposed based on statistical physics arguments. Among other approximate methods, let us mention the adaptive cluster expansion \cite{Cocco2011} which controls the accuracy of the approximation at a cost of a higher computational complexity involving computation of entropies of growing clusters, and the probabilistic flow method \cite{sohl2011new} introducing a relaxation dynamics to certain trial distribution. However, both schemes remain computationally expensive and thus not suitable for large systems, and rely on fine tuning of auxiliary parameters. An alternative method which uses the full information contained in the samples, has been suggested and rigorously analyzed in \cite{ravikumar2010high}. Although it has been shown in \cite{montanari2009graphical} that this estimator is unable to correctly reproduce the underlying graph of the original model at low temperatures, until lately with certain modifications it remained the state-of-the-art practical method \cite{aurell2012inverse,ekeberg2013improved,Decelle2014}. Partly anticipating on our results, we show later in this paper that this \emph{Regularized Pseudo-Likelihood Estimator} can be turned into an exact and universal method if completed with a rather natural, but key, ingredient: a post-inference thresholding of reconstructed couplings.

The problem of designing a universal learning algorithm with polynomial computational complexity \cite{montanari2015computational} that achieves exact graph topology reconstruction for arbitrary Ising models in all regimes was resolved only recently in \cite{bresler2008reconstruction,Bresler2015}. The biggest challenges addressed were the low temperature regime and long-range correlations, which are known to be particularly difficult for learning. Nonetheless, the computational cost of these algorithms is still high, and scales as a polynomial of high degree in the number of nodes \cite{bresler2008reconstruction}, or double exponential in the maximum node-degree $d_{\max}$ and in the maximum interaction strength \cite{Bresler2015}. Moreover, both algorithms require prior information on the bounds on the interaction strengths, i.e. positive $\alpha$ and $\beta$ such that $\alpha \leq \vert J_{ij} \vert \leq \beta$ for all $(i,j) \in E$, as well as the knowledge of $d_{\max}$.

In an attempt to determine the \emph{optimal} number of samples needed for reconstructing the graph, information-theoretic bounds were derived in \cite{santhanam2012information}. We emphasize three salient features of these bounds. First, the \emph{optimal} number of samples $M_{opt}$ for perfect graph recovery scales exponentially with the maximum interaction value and node-degree, $M_{opt} \propto e^{c\gamma}$, where $\gamma = \beta d_{\max} + h_{\max}$ and $h_{\max}$ denotes an upper bound on the absolute values of magnetic fields. Although it was shown that $c \in [1,4]$, the precise value of $c$ remains unknown; in this manuscript, we refer to this range of $c$ as to the \emph{optimal regime} with respect to the dependence of the number of samples on $\gamma$. Intuitively, this exponential scaling requirement can be attributed to the typical waiting time for collecting sufficient number of ``non-trivial'' samples, i.e., those that are different from the ground states. This waiting time is more pronounced in the low temperature regime when $\gamma$ is large. Second, for finite $d_{\max}$ the dependence on the number of variables $N$ is very weak: $M_{opt} \propto \ln N$. This logarithmic dependence represents the amount of information needed for hypothesis testing over the set of $C^{d_{\max}}_{N}$ candidate neighborhoods of a given vertex \cite{yu1997assouad}. Third, the number of required configurations grows as $\alpha$ decreases, since it is difficult to distinguish between the presence of a very weak coupling and its absence. In particular, in the limit of small $\alpha$, $M_{opt} \propto 1/\alpha^{2}$.

In what follows, we discuss two exact methods for solving the inverse Ising problem. The first method is based on the Regularized Pseudo-Likelihood Estimator of \cite{ravikumar2010high} supplemented with a post-optimization parameter thresholding procedure. We prove that this ingredient makes this estimator exact, meaning that the algorithm can reconstruct an arbitrary Ising model with an appropriate number of samples. The second algorithm that we introduce is an exact estimator based on the Interaction Screening method. By setting up a framework for an empirical assessment of the performance of the algorithms guided by the information-theoretic arguments presented above, we show that our new estimator outperforms the Pseudo-Likelihood based algorithm, and requires in all test cases a number of samples lying within the information-theoretic \emph{optimal regime}.

\section*{Results}

\subsection*{Regularized pseudo-likelihood estimator}

A widely used approach aiming at achieving the optimal scalings was suggested in \cite{ravikumar2010high}, where estimation of model parameters is performed based on the so-called pseudo-likelihood acting as a surrogate for the intractable log-likelihood function. The method is based on maximizing a set of local \emph{Regularized Pseudo-Likelihood Estimators} (RPLE). Each of them can be interpreted as a regularized probability of a single spin conditioned on the remaining $N-1$ spins in the system given by
\begin{equation}
\mathcal{L}_{i}(J_i,H_i) = \left\langle \ln \frac{1}{1+e^{-2\sigma_i(H_i+\sum_{j\neq i}J_{ij}\sigma_j)}} \right\rangle_M
\hspace{-0.2cm} -\lambda\Vert J_{i} \Vert_{1}, \label{eq:PL}
\end{equation}
where $\langle f(\sigma)\rangle_M = M^{-1}\sum_{m=1}^{M} f(\sigma^{(m)})$ is the notation for the empirical average; $J_i$ and $H_i$ are the optimization parameters; and $J_i$ is the shorthand notation for $\{J_{ij}\}_{j \neq i}$. The sparsity promoting $\ell_{1}$ regularization term $\Vert J_{i} \Vert_{1} = \sum_{j \neq i} \vert J_{ij} \vert$ is important as it discourages the minimizer $\widehat{J}_i$ from being dense by effectively pushing the interaction values toward zero whenever an edge is absent. In the original version of the algorithm, the graph structure is identified as a set of edges carrying couplings that were not set to zero by the RPLE. Guarantees for perfect graph reconstruction following this procedure rely on a rather restrictive set of conditions that are not always satisfied and are hard to verify in practice \cite{ravikumar2010high}. Models known to satisfy these conditions are particular ferromagnetic models at high temperature, but this procedure provably fails in other regimes, most noticeably at low temperatures \cite{montanari2009graphical}. A natural extension of this algorithm which uses a post-estimation thresholding of a part of non-zero couplings was introduced in \cite{aurell2012inverse}. In this scheme, all recovered $J_{ij}$ satisfying $|J_{ij}|<\delta$, where $\delta$ is a chosen threshold, are declared to be zero. However, the performance of the RPLE-based algorithm with thresholding has never been rigorously analyzed, and until now it was believed that any RPLE scheme fails in the low temperature regime, following theoretical indications \cite{montanari2009graphical} and experimental studies conducted in a framework that does not fully account for the sample complexity structure of the inverse Ising problem \cite{aurell2012inverse}. The reason why previous numerical studies were showing a failure of RPLE with thresholding at low temperatures is most certainly due to the hidden dependence of the required number of samples $M^{*}$ on the strength of the couplings (inverse temperature $\beta$) in the original analysis \cite{ravikumar2010high}, which resulted in tests of the reconstruction quality as a function of inverse temperature assuming a constant number of samples only. At the same time, it is clear that in the low-temperature regime the Boltzmann probability measure concentrates on the ground state samples, i.e. most of the samples in a typical batch would correspond to less-informative ground state configurations. Hence, an assessment of empirical performance should be based on a setting where the number of provided samples is exponentially increasing, in agreement with information-theoretic dependencies \cite{santhanam2012information}. We take this fact into account in the numerical experiments presented below.

In the Supplementary Text, we prove that there exists a minimum number of samples $M^{*}$ for which the error on the estimated couplings is bounded by $\alpha/2$, so that choosing $\delta=\alpha/2$ indeed leads to a perfect reconstruction of the graph topology. Hence, our first result states that the RPLE with a post-evaluation thresholding is exact: in the worst case, the required number of samples scales at most as $M^{*} \propto \exp(8\gamma) \ln N / \alpha^{2}$, see Supplementary Text for details. Note that the parameter estimation problem for each vertex is independent, and the optimization can be carried out separately for each spin. As we explain below, the symmetrized estimate of coupling associated with the edge $(i,j)$ is obtained as an average of local estimates $(\widehat{J}_{ij} + \widehat{J}_{ji})/2$. This parallelization of local reconstructions is lost when the optimization is performed globally over the entire graph \cite{Decelle2014}.

\begin{figure}[!htb]
\begin{center}
\includegraphics[width=\columnwidth]{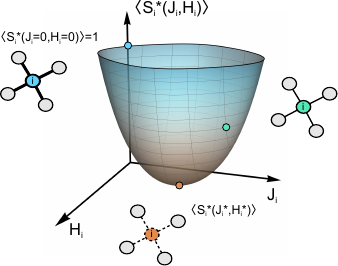}
\caption{{\bf Interaction Screening Objective for different probe values of model parameters in the large $M$ limit.}
The ISO is an empirical average of the inverse of the factors in the Gibbs measure, and its screening property becomes apparent in the limit of large number of samples. Changing the value of the probing parameters $(J_i,H_i)$ in the ISO alters the effective interaction strength of $\sigma_i$ with its neighbors. This mechanism is schematically represented in the figure where the value of ISO for different values of probing parameters is depicted. When the probing parameters are equal to the true ones $(J^*_i, H^*_i)$, the ISO completely screens this interaction making $\sigma_i$ effectively independent of its neighbors. With some analysis, this can be shown to be equivalent to the ISO attaining its minimum at the true parameters of the model.}
\label{fig:Figure1}
\end{center}
\end{figure}

\subsection*{Interaction Screening method}

Recently, we introduced the first exact reconstruction algorithm having the same parametric dependence as the information-theoretic bound and termed the \emph{Regularized Interaction Screening Estimator} (RISE) \cite{screening2016theoretical}. Our theoretical analysis showed that the RISE has a lower theoretical sample complexity for perfect graph recovery, compared to the one derived here for the RPLE with thresholding, guaranteeing that a number of samples $M^{*} \propto \exp(6\gamma) \ln N / \alpha^{2}$ is sufficient for reconstruction of the graph structure, see the Supplementary Text and \cite{screening2016theoretical} for details. But the factors $6$ and $8$ in the exponents of the RISE and the RPLE, respectively, are likely to be an artifact of the employed proof techniques and is not tight as indicated by the computational experiments in this paper.

The RISE is based on the minimization of the Interaction Screening Objective (ISO)
\begin{equation}
S_i(J_i,H_i) = \langle \exp(-\sum_{j\neq i} J_{ij} \sigma_{i} \sigma_{j} -H_{i}\sigma_{i}) \rangle_M
\label{eq:Estimator}
\end{equation}
over the probe vector of couplings $J_{i}$ and the probe magnetic field $H_{i}$ for a given spin $i$. The ISO, as its name suggests, is constructed based on the property of ``interaction screening" which is illustrated in Fig.~1. As a consequence of this property, in the limit of large number of samples the unique minimizer of the convex ISO objective is achieved at $(J_i, H_i) = (J^*_i, H^*_i)$. A simple derivation of this fact is presented in the Materials and Methods section. In the RISE construction, the ISO is appended with the $\ell_1$ regularizer in order to promote sparsity \cite{screening2016theoretical}. In this paper, we introduce a modification to the RISE that leads to a new exact learning method for the inverse Ising problem, that we call the logRISE and which takes the following form:
\begin{equation}
(\widehat{J}_i,\widehat{H}_i) = \argmin_{( J_i,H_i)} \Big[ \ln S_{i}(J_i,H_i) + \lambda \Vert J_i \Vert_{1} \Big].
\label{eq:RISE}
\end{equation}

The name logRISE comes from the fact that instead of the ISO itself, we use its logarithm to form the logRISE objective \eqref{eq:RISE}. Obviously, in the absence of the regularizer (for $\lambda = 0$), taking the logarithm of the ISO does not change its minimizer. However, this difference is crucial for non-zero values of the regularization term, which suggests that logRISE might have good properties for the reconstruction problem due to a particular form of its first and second derivatives (see the Supplementary Text for additional explanations and details).

Unfortunately, the proof techniques used for deriving bounds on scaling for the RPLE and the RISE provide less tight expressions when applied to the estimator logRISE, since it no longer can be represented in a form of finite functional sum over individual samples. Our analysis states that the number of required samples for logRISE in the worst case scales as $M^{*} \propto \exp(10\gamma) \ln N / \alpha^{2}$ for guaranteeing the reconstruction the structure of the underlying Ising model with high probability. Given looseness of the theoretical analysis in this case, the empirical assessment of the performance of thelogRISE and its comparison with the RPLE is required. We provide a detailed numerical study of the quality of different estimators below.

As we show through a rigorous analysis in the Supplementary Text, the $\ell_{1}$ regularizer plays an important role for all of the estimators since it reduces the required sample complexity for perfect topology reconstruction from quasi-linear to logarithmic in the number of spins $N$. However, the performance of the RPLE, the RISE and the logRISE, and hence the number of required samples $M^{*}$ is dependent on the regularization coefficient $\lambda$. The choice of $\lambda$ needs to account for the following tradeoff: if $\lambda$ is too small, the estimation is prone to noise; and if $\lambda$ is too large, it introduces a bias in the estimated couplings toward zero. The optimal value of $\lambda$ is unknown \emph{a priori}. In the Supplementary Text we present detailed simulations for different topologies which show that for achieving correct graph reconstruction with probability $1-\epsilon$, the choice $\lambda=c_{\lambda}\sqrt{\ln(N^2/\epsilon)/M}$ is appropriate when no additional information about the model is available, with $c_{\lambda}\simeq 0.2$ for the RPLE, $c_{\lambda}\simeq 0.4$ for the RISE and $c_{\lambda}\simeq 0.8$ for the logRISE. We use these values for $\lambda$ in all numerical experiments reported below. Given a sufficient number of samples, other techniques such as consistency cross-validation can be employed for selecting the optimal value of the regularization coefficient on a case by case basis. An illustration of this approach alongside some practical remarks are provided in the Supplementary Text.

\begin{figure}[!htb]
\begin{center}
\includegraphics[width=\columnwidth]{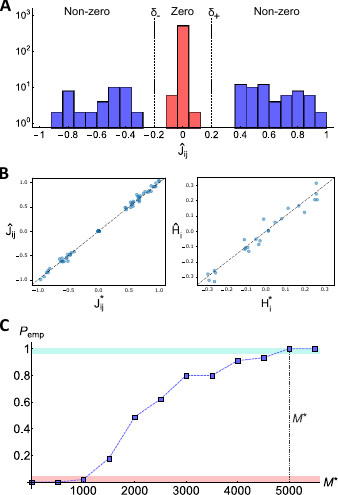}
\caption{{\bf Reconstruction of the graph topology and the values of parameters with the logRISE.} Reconstruction procedure for an Erd{\"o}s-R{\'e}nyi graph with $N=25$ and average degree $\langle d\rangle = 4$ given $M=5000$ configurations. The couplings and magnetic fields are generated uniformly at random in the range $[-1.0,-0.4]\cup [0.4,0.1]$ and $[-0.3, 0.3]$, respectively. \textbf{A:} The symmetrized estimate of coupling $\widehat{J}_{ij}$ associated with the edge $(i,j)$ is obtained as an average of local estimates $(\widehat{J}_{ij} + \widehat{J}_{ji})/2$. When the regularizing parameter $\lambda$ is appropriately chosen, and the number of samples is sufficient, gaps emerge in the estimated couplings $\widehat{J}_{ij}$ around $\delta_{+}>0$ and $\delta_{-}<0$, separating the estimated couplings that are close to zero and those with higher intensities in absolute value. The values below the threshold are then set to zero to obtain an estimate of the graph structure. \textbf{B:} Once the graph structure is learned, the parameters are re-estimated by optimizing the unregularized objective only over the edges in the reconstructed graph. The reduction in the number of free optimization variables from $N$ to $d_{max}+1$ greatly improves the estimates. The resulting values are shown in the scatter plot. \textbf{C:} Empirical probability of successful structure recovery $P_{\text{emp}}$ over $L=45$ independent runs as a function of the number of samples $M$. For the logRISE, the smallest number of samples for which $P_{\text{emp}}=1$ is given by $M^{*} = 5000$.}
\label{fig:Figure2}
\end{center}
\end{figure}

\subsection*{Learning structure and parameters of the model}

We state our three-step algorithm for learning the underlying graph and the parameter values of the Ising model using the RPLE or the logRISE (the same algorithm applies to the RISE). First, given $M$ samples, we find the minimizer of the objective \eqref{eq:PL} or \eqref{eq:RISE}, respectively, at each node $i \in V$, and obtain a collection of estimated parameters $(\widehat{J}_i,\widehat{H}_i)$. Given that both estimators are convex, any appropriate convex optimization method can be used to find the minimizer of the objective function, the simplest one being a plain gradient descent supplemented with an additional projection step due to non-differentiability of the $\ell_1$ regularization term. For our numerical experiments, we used the Ipopt optimization software \cite{biegler2009large}, however, as we comment in the Supplementary Text, better choices such as composite-type gradient descent methods exist for experiments with very large networks \cite{nesterov2007gradient, agarwal2010fast}.

Given a sufficient number of samples $M$, a typical histogram of couplings estimated by the RPLE, the RISE or the logRISE takes the form shown in Fig.~2~(A). Notice the emergence of gaps separating a group of inferred couplings that are close to zero from those with significantly bigger intensities in absolute value. In the second step, we threshold the inferred couplings below the observed gaps to zero. The edges associated with the remaining non-zero couplings form the reconstructed graph $\widehat{G}$. Finally, we optimize the unregularized objective for each of the three estimators, i.e. setting $\lambda=0$, but only over the couplings corresponding to the edges in $\widehat{G}$, and obtain our final estimates $(\widehat{J}_i,\widehat{H}_i)$. This procedure is illustrated in Fig.~2~(B) for the logRISE on an Erd{\"o}s-R{\'e}nyi graph with $N=25$ nodes and spin glass couplings, where the scatter plot of predicted versus true values of the model parameters is presented, and only parameters over the already reconstructed graph from $M = M^{*} = 5000$ samples have been accounted for. We see that even using a small number of samples, in this example the minimal amount for a correct structure recovery, the numerical values of the parameters are also reconstructed with a very good accuracy that increases when more samples are provided.

\begin{figure}[!htb]
\begin{center}
\includegraphics[width=\columnwidth]{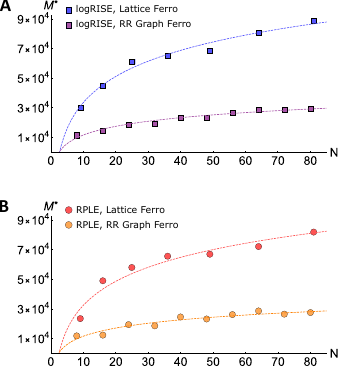}
\caption{{\bf Verification of the logarithmic scaling with the size of the system.} Scaling of $M^{*}$ with the number of spins $N$ for (\textbf{A}) the logRISE and (\textbf{B}) the RPLE  obtained using samples produced in the cases of the ferromagnetic Ising model over a double periodic two dimensional lattice with $\beta=0.7$ and ferromagnetic random regular graphs with degree $d=3$ for $\beta=1.0$. In all cases, we observe a logarithmic growth of $M^*$ with respect to $N$ which is in agreement with the information-theoretic bounds, as well as our theoretical analysis for the estimators.}
\label{fig:Figure3}
\end{center}
\end{figure}

In order to have statistical confidence in our results, we determine $M^{*}$ as follows. Progressively increasing values of $M$, the reconstruction experiment runs $L$ times, using $L$ independent sets of $M$ samples. Based on the number of successful topology reconstructions $L_{\text{succ}}$, one can define the empirical probability of reconstruction $P_{\text{emp}} = L_{\text{succ}}/L$. We define $M^{*}$ as the minimum $M$ for which $P_{\text{emp}}=1$, see Fig.~2~(C) for a typical example. The value of $L$ that we use in our computations comes from the requirement of a perfect topology reconstruction with probability greater than $1-\epsilon$, where we fix $\epsilon=0.05$. In other words, it is essential to get $L=45$ successful reconstructions in a row in order to make sure that the probability of correct topology recovery is above $0.95$ with confidence at least $90\%$, as we explain in the Supplementary Text. We use this value of $L$ in the computations throughout the text.

We performed extensive numerical experiments to obtain empirically the minimal number of samples $M^{*}$ required for perfect graph reconstruction for different topologies and types of interactions. We carried out the numerical experiments for all of the three estimators considered in this paper. However, for the sake of simplicity and for the clarity of presentation, in what follows in the main text we present numerical results only for the logRISE, which is the central object of the present study, and for the RPLE, which is the state-of-the-art method for the inverse Ising problem. Note that throughout the manuscript we present comparisons of the logRISE with the exact and universal version of the RPLE, i.e. corrected through our thresholding procedure. The corresponding scalings for the RISE are available in the Supplementary Text.

\begin{figure*}[!htb]
\begin{center}
\includegraphics[width=1.92\columnwidth]{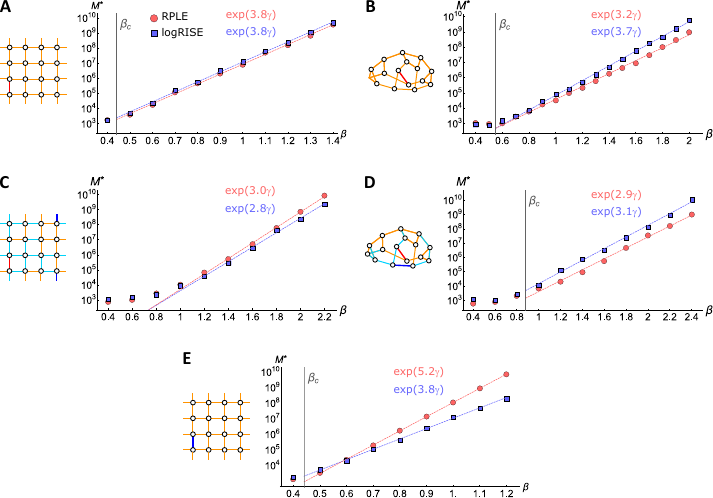}
\caption{{\bf Scaling of $M^{*}$ with the couplings strength.} Comparison of the performance of the logRISE and the RPLE is presented for five different ensembles of Ising models. Due to a weak dependence $M^{*} \propto \ln N$, we consider graphs of size $N=16$ which allowed us to produce independent samples through an exhaustive enumeration of spin configurations. The first four cases correspond to {\bf (A)} a ferromagnet on a square lattice with double-periodic boundary conditions, {\bf (B)} a ferromagnet on a 3-regular random graph,  {\bf (C)} a spin glass on a periodic lattice and {\bf (D)} a spin glass on a 3-regular random graph. The most difficult reconstruction test case for both algorithms, a ferromagnetic lattice with a weak anti-ferromagnetic impurity, is presented in {\bf (E)}. The phase transition points in the corresponding infinite systems are indicated as $\beta_{c}$. An exact pictorial representation of the corresponding Ising model is portrayed on the left-hand side of each. Ferromagnetic couplings equal to $\beta$ and $\alpha=0.4$ are colored in orange and red respectively. Anti-ferromagnetic couplings equal to $-\beta$ and $-\alpha$ are respectively colored in turquoise and blue.}
\label{fig:Figure4}
\end{center}
\end{figure*}

We first verify the logarithmic scaling of $M^{*}$, claimed in our theoretical analysis for RPLE and logRISE, with respect to the number of spins $N$ in ferromagnetic Ising models without magnetic fields ($J^{*}_{ij} > 0$, $H^{*}_{i} = 0$), defined on two topologies: square lattice with periodic boundary conditions and random 3-regular (RR) graphs. The choice of the ferromagnetic models has been dictated by the need to generate independent samples for large values of $N$, and given that for spin glass models this is a non-trivial task \cite{mezard2009information}. For the two aforementioned topologies we generate independent samples using Glauber dynamics for different values of $N$ in the low-temperature regime where the correlations are long-range: we have used $J^{*}_{ij}=0.7$ for the lattice ensemble and $J^{*}_{ij}=1.0$ for the RR graphs ensemble. The minimal required sample size $M^{*}$ on both topologies are presented in Fig.~3. We see that $M^{*}$ exhibits a logarithmic dependence on $N$ for both estimators, the logRISE and the RPLE.

The major difference in performance between the estimators is observed in the scaling with respect to $\gamma = \beta d_\text{max} + h_\text{max}$. This is critical since a favorable exponent allows the algorithm to have a lower sample complexity in the low-temperature regime where known algorithms either do not work or exhibit poor scaling. An extensive numerical study is presented in Fig.~4, where we study quasi-homogeneous systems with ferromagnetic type couplings (A, B, and E) and spin glass type couplings (C and D) on two topologies: square lattice with a double-periodic boundary conditions (A, C, and E) and random regular graphs (B and D). This choice of topologies eliminates fluctuations with respect to the heterogeneity of node degrees, so that it becomes easier to extract the right scaling with respect to $\beta$ and $d$. In order to disentangle the effects of $\alpha$ and $\beta$, we always fix one (for ferromagnets) or two (for spin glass systems) couplings to $\alpha$ or $-\alpha$, which is different from the interaction values $\pm \beta$ carried by the rest of the edges. Therefore, $\beta$ can be conveniently thought of as the inverse temperature of the model. In order to investigate the effect of temperature on the scalings, we deliberately set magnetic fields to zero, and fix the thresholding parameter to $\delta = \alpha/2$. The test cases (A, B, C, and D) represented in Fig.~4 show that overall the RPLE and the logRISE demonstrate similar scaling properties. Notice that there exists a qualitative difference in the scaling behavior between the low and high temperature regimes, with an exponential scaling for both estimators observed for large $\beta$. Our numerical study shows that from the learning perspective, the ferromagnetic model on the two-dimensional lattice appears to be the most challenging class of Ising models for both the logRISE and the RPLE. It has the highest scaling exponent with respect to $\gamma$ and hence the largest sample complexity for the inverse Ising problem. This observation supports theoretical evidence that this case is among the hardest class of models for learning \cite{tandon2014information}. In particular, this finding shows that, paradoxically, the inverse Ising problem on a planar ferromagnetic model is harder to learn compared to spin glass models while the direct problem of drawing independent samples from the former can be incomparably easier than from the latter \cite{mezard2009information}.

The ultimately hardest case for the reconstruction problem is unknown. However, we were able to construct a slight variant of the ferromagnetic model on a lattice that appears to be even harder for all algorithms considered: a ferromagnetic model with a weak anti-ferromagnetic interaction, i.e. an edge carrying the coupling $-\alpha$. In the Discussion section, we present intuitive arguments why this case should be fundamentally hard. The results for the extraction of the $M^{*}$ in this model instance are presented in Fig.~4~(E). We see that the logRISE has a strikingly better scaling exponent compared to the RPLE. Remarkably, in this test case the scaling exponent of the RPLE is significantly larger than the information-theoretic upper bound, wile the corresponding value for the logRISE lies within the \emph{optimal regime} in terms of the information-theoretic predictions. We summarize the scaling behavior of the estimators in the Discussion section.

\subsection*{Application to a real system: D-Wave quantum computer}

In order to evaluate the performance and robustness of the estimators in a non-synthetic case, we apply the logRISE and the RPLE to real data produced by the D-Wave 2X quantum annealer ``Ising'' at Los Alamos National Laboratory. The D-Wave computer \cite{bunyk2014architectural} has been designed for solving binary quadratic optimization problems in the form of Ising models through quantum annealing, i.e. slowly transforming an initially prepared state of the system to the ground state of the desired input Ising Hamiltonian encoded on its chip. Because of the thermal noise in the system, a single annealing run may end in one of the excited states instead of the desired ground state. In practice, the device attempts to find the target ground state by re-running the annealing multiple times, and producing as output the best solution found. Previous experiments with D-Wave report that the produced samples are distributed according to the Boltzmann distribution at some effective temperature \cite{benedetti2016estimation}, related but not equal to the native temperature at which D-Wave operates. This effective temperature is naturally low as D-Wave contains superconducting elements as a part of its architecture. Due to the temperature rescaling effect, as well as inevitable biases present in this analogue device, the effective Ising model from which the samples are produced does not exactly correspond to the input Ising model. It then becomes interesting to see how the structure of the distorted effective Ising model is related to the one encoded in the chip. This task is exactly what the methods presented in our paper are designed to solve, making it a good real-world application for testing their performance.

Let us describe the procedure that we followed for generating the data. Our goal was to check the performance of the algorithms on a noisy heterogeneous instance, both in node degrees and couplings as well as magnetic fields. Hence, we encoded an Ising model with random couplings and magnetic fields, distributed uniformly in the range $[-0.16,-0.02]\cup [0.02,0.16]$. We also chose to encode these couplings in a region of the chip with the highest concentration of broken qubits that are inevitably present and can potentially create additional noise. The topology of this portion of the chip containing $N=62$ qubits is illustrated in gray in Fig.~5~(A). We observed that the initial Ising model got distorted while being implemented on the chip. From several trial tests, we inferred that the effective rescaling factor in this regime roughly fluctuates around $\beta_{\text{eff}} \approx 12$, although this factor is different for individual model parameters. Since the precise values of the couplings and magnetic fields actually implemented on the chip are unknown, the only ``ground-truth" available to us in this experiment is the topology of the portion of the chip we encoded our model on. However, let us point out that due to the complexity of the D-Wave architecture and a possible conflict of superconducting loops representing couplers between qubits, it is \emph{a priori} unclear whether the resulting topology of the effective Ising model will necessarily remain unchanged.

The maximum number of annealing runs for a given Ising model implementation is limited to $10^4$ by standard system settings on the D-Wave. We collected $5\times10^5$ samples corresponding to the same input model specified above by obtaining $50$ batches of $10^4$ samples each, and provided them as an input to the logRISE and the RPLE. Notice that each additional implementation of the same chosen Ising model for each batch in principle corresponds to a different actual Ising Hamiltonian owing to a different concrete realization of random biases; this creates an additional source of noise in our data. The reconstructed model parameters are presented in Fig.~5~(A) and Fig.~5~(B). We emphasize that it is difficult to disentangle the effects of statistical errors due to the finiteness of the number of samples, and the errors due to noise.

\begin{figure*}[!htb]
\begin{center}
\includegraphics[width=1.92\columnwidth]{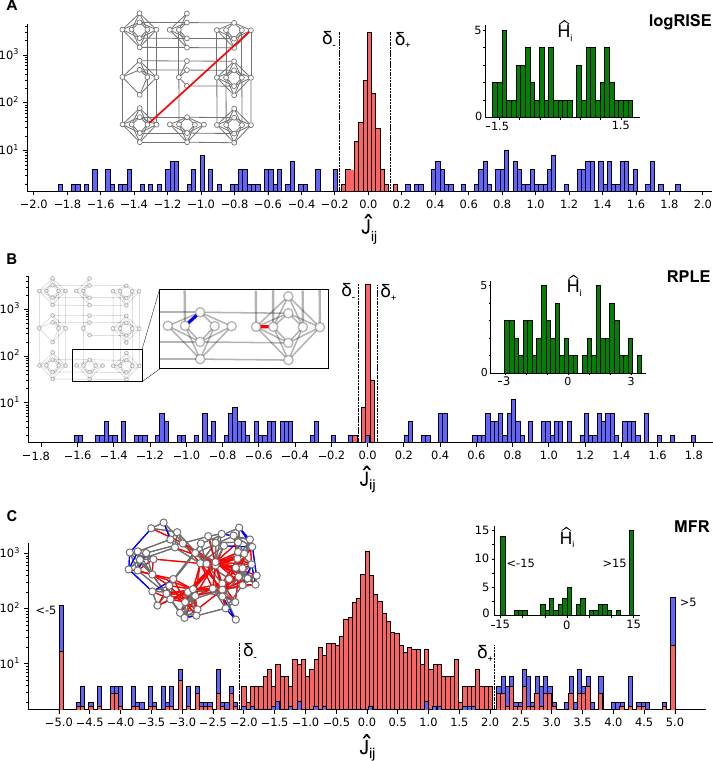}
\caption{{\bf Reconstruction of the structure of a portion of the D-Wave annealer chip using $5\times10^5$ samples}. This part contains 62 qubits with heterogeneous connectivity, couplings and magnetic fields. Reconstructed couplings are presented for {\bf (A)} the logRISE, {\bf (B)} the RPLE and {\bf (C)} the Mean-Field Regime (MFR) of the RPLE and the logRISE. On each histogram in the main plots, bars corresponding to the edges actually present on the chip are colored in blue, while non-existent connections are colored in red. The reconstructed magnetic fields are shown in green in a separate histogram on the right-hand side. A topology of the reconstructed structure is depicted on the left-hand side, with correctly reconstructed edges, missing edges and incorrectly reconstructed edges colored in gray, blue and red respectively. While the MFR exhibits a poor behavior, as expected at such low temperatures, both the logRISE and the RPLE are achieving similarly good performance. Notice that wheras there exists a thresholding procedure that produces a perfect network recovery with the logRISE, it is not the case for the RPLE as one existing coupling has been set to zero.}
\label{fig:Figure5}
\end{center}
\end{figure*}

For structure reconstruction, we chose to threshold the parameters $J_{ij}$ in the tail of a set of couplings reconstructed in the vicinity of zero. Given this choice of threshold, we found that both algorithms are quite robust to noise and are able to accurately reconstruct the graph topology, making only a few false positives and negatives. The reconstructed topologies are shown in the left of Fig.~5~(A) and Fig.~5~(B). Notice that although the RPLE makes local errors, detecting one false postive and one false negative connections between neighboring spins, the logRISE misclassifies a non-existing edge as existing in a clearly non-local fashion, meaning that the vertices it misclassifies as neighbors are far away in graph theoretic distance on the D-Wave chip. Interestingly, while in the case of the logRISE it is possible to choose an optimal threshold that allows one to completely separate zero couplings from non-zero ones and thus reconstruct the structure of the chip perfectly, no such thresholding is possible for the result produced by the RPLE, suggesting that the RPLE needs more samples before this separation becomes possible. Finally, notice that according to the histograms on reconstructed magnetic fields in the right insets of Fig.~5~(A) and Fig.~5~(B), the RPLE seems to make larger errors in the reconstruction of magnetic fields that should be of the same order as couplings according to our input Hamiltonian.    

As we pointed out in the Introduction, a plethora of other methods have been proposed for the inverse Ising problem, but the majority of them are either too computationally expensive for practical applications, or fail at low temperatures, sometimes even when an infinite number of samples is provided. To illustrate the value of exact algorithms, especially for problems at low temperatures (such as this application), we compare the results obtained from the logRISE and the RPLE to those from Mean-Field type methods, see Fig.~5~(C). The particular scheme that we used for comparison is obtained from a high-temperature expansion of our estimators, and is closely related to the na\"{i}ve mean-field method of statistical physics which performs well at high temperatures. See Methods and Materials section for a detailed description of the method and related discussions. As expected for such systems with strong and long-range correlations, this method utilizing only information contained in magnetization and pairwise correlations behaves poorly, incurring a very large number of false positives and false negatives. This illustrates an importance of taking into account higher-order interaction in data samples for a reliable reconstruction in the low-temperature regime.

\begin{figure*}[!htb]
\begin{center}
\includegraphics[width=1.92\columnwidth]{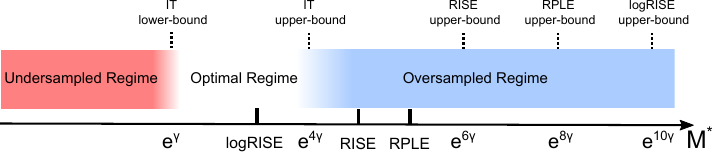}
\caption{{\bf Theoretical and empirical worst-case scaling of $M^{*}$ with respect to $\gamma$.} This figure summarizes the main theoretical and empirical results of this paper for the inverse Ising problem. The red region represents the undersampled regime where the number of samples is insufficient for perfect graph reconstruction from the information theory perspective. The existence of an exact algorithm, albeit with an exponential computational complexity, has been proven for $M \propto e^{4 \gamma}$, and thus represents an upper bound on the optimal number of samples $M_{\text{opt}}$ that must lie in the white region, named the \emph{optimal regime}. The quantities $e^{6 \gamma}$ and $e^{8 \gamma}$ and $e^{10 \gamma}$ denote our theoretical upper bounds on the scaling for the RISE, the RPLE and the logRISE, respectively. However, these bounds are not tight, and the worst-case empirical scalings observed in our numerical experiments were much lower; these values are indicated in the chart as ``RISE'', ``RPLE'', and ``logRISE" and correspond to  $e^{4.5 \gamma}$, $e^{5.2 \gamma}$ and $e^{3.8 \gamma}$, respectively (see the Supplementary Text for additional details on the scaling of the RISE). Remarkably, the empirical scaling for the logRISE lies within the \emph{optimal regime}.}
\label{fig:Figure6}
\end{center}
\end{figure*}

\section*{Discussion}

In Fig.~6, we comparatively present the algorithmic scalings that summarize the main theoretical and empirical results of our paper. All of the three considered estimators for the inverse Ising problem have a better worst-case empirical scaling compared to their theoretical estimates. Remarkably, the empirical sample complexity of the the logRISE algorithm introduced in this paper lies in the \emph{optimal regime} with respect to the information-theoretic predictions, outperforming all existing methods. The worst-case scalings are based on the hardest case for the learning problem that we were able to construct. To describe the logic behind this case, we first mention observations in existing literature and then provide intuitive arguments regarding the way the structure of the underlying graph and the nature of interactions affects the hardness of reconstruction. There are strong theoretical indications that ferromagnetic-type spin-systems are among the models requiring maximum number of samples to be learned. Information-theoretic bounds suggest that these models are at least as hard to learn as any other model \cite{tandon2014information}. Moreover, the presence of strong long-range correlations is known to be a challenging situation to deal with \cite{Bresler2015}. This hardness of learning ferromagnetic models is consistent with our numerical studies in which ferromagnetic random graphs and especially ferromagnetic lattices are the cases requiring the largest amount of samples. Intuitive explanations for this behavior are twofold. As mentioned earlier, ferromagnetic models are more prompt to develop strong long-range correlations at low temperatures, especially on lattices, and they tend to favor two configurations that are the ground states. Long-range correlations make it less likely to obtain non-trivial samples, i.e. fluctuations around ground states, that are crucial to obtain information about the detailed structure of the graph that is crucial for the reconstruction. This translates into a need for a larger number of samples, proportional to the likeliness of such fluctuations which is typically exponentially suppressed in $\gamma$. Moreover, when several similar models share identical ground states, it becomes very hard to make a distinction between them solely based on configurations close to their ground states. This mechanism can be illustrated very simply using an extreme example of three spins with homogeneous couplings forming a chain that is either open or closed, forming a triangle. Deciding which chain is formed is impossible for a ferromagnetic system when only the ground states $\pm(1,1,1)$ are observed. However it is an easy task for an anti-ferromagnetic system as an open chain has two ground states $\pm(1,-1,1)$ whereas the close chain has six ground states $\pm(1,1,-1)$, $\pm(1,-1,1)$, $\pm(-1,1,1)$.

The hardest test case studied in our numerical experiments contains an extra ingredient that makes the inverse Ising problem even more challenging: an additional weak negative coupling or ``anti-ferromagnetic impurity" added on top of the ferromagnetic model on a lattice. This weak anti-ferromagnetic bond has the effect of weakening or cancelling the correlation between the two spins that it connects. Consequently, it becomes difficult to distinguish between the presence of this weak negative coupling from its absence. Although we do not claim with certainty that this model is the hardest to learn, we believe that any such difficult-to-learn model is likely to include the features outlined above.

We proved that the three techniques explored in this paper, the logRISE, the RISE and the RPLE, are exact and universal methods to solve the inverse Ising problem. Exactness and universality in this context mean that these methods reconstruct couplings and magnetic fields up to any given accuracy with a sufficient but finite number of samples and for every Ising model regardless of its structure, density, temperature or any other property that characterize it. While in the present article we focused on the quantification of the scaling of the number of required samples with structural properties and temperature of \emph{sparse} systems, it remains an interesting question left to exploration for \emph{dense} models, for instance of the Curie-Weiss or Sherrington-Kirkpatrick type \cite{mezard2009information}. In these models, the exponential scaling with coupling intensities and degrees, denoted by $\gamma$ for sparse models, will be more intricate. It seems reasonable to expect that the sample requirement scales exponentially with the typical ``energy per spin". For instance, in the Curie-Weiss type models with all $J_{ij}\geq 0$, this quantity is $\gamma_{\text{CW}} \approx  h_i + \frac{1}{N}\sum_{j \neq i}J_{ij}$, while in the Sherrington-Kirkpatrick type models where $J_{ij}$ are centered random variables it reads $\gamma_{\text{SK}} \approx  h_i + \frac{1}{\sqrt{N}}\sum_{j \neq i}\sqrt{\mathrm{Var}(J_{ij})}$. We also note that in these dense models there is no longer any reason to expect that the sample complexity requirement scales logarithmically with the system size \cite{santhanam2012information}, instead we expect it to exhibit a polynomial dependence. Note that in this case the inclusion of the $\ell_1$ regularizer in the logRISE and the RPLE is no longer necessary since there is no sparsity pattern to promote.

In conclusion, in this paper we showed both theoretically and experimentally that an arbitrary Ising model can be reconstructed exactly with a information-theoretically minimum number of samples using the introduced Interaction Screening method. Additionally, no prior knowledge on the graph and associated parameters is required to implement the algorithm, making it a very practical choice for applications. The practical advantages of our methods have been illustrated on a real data coming from a D-Wave quantum computer. We also provided a sample complexity analysis of the popular Regularized Pseudo-Likelihood Estimator showing the logarithmic scaling in system size for arbitrary Ising models, albeit with a higher worse-case scaling with respect to the inverse temperature when compared to the logRISE. We demonstrated the paradoxical relation between sampling and learning, showing that the instances that are easier for one task are harder for the other. In the Materials and Methods section, we point out a curious connection to the mean-field approximation at high temperatures. Interestingly, the second-order high-temperature expansion of all exact estimators considered in this paper provides an identical reconstruction scheme, valid in the limit of weak couplings. This high temperature regime is related to learning methods based on the the well-known na\"{i}ve mean-field approximation in statistical physics. Finally, even though this paper is dedicated to the reconstruction of Ising models, the Interaction Screening method can be generalized to graphical models with higher-order interactions and non-binary alphabets, including those described by Hamiltonians over continuous variables. Exploration of these research directions is underway.

\section*{Materials and Methods}

\subsection*{Interaction Screening property}

Here, we present present a simple argument that illustrates the fact that in the limit of large number of samples the unique minimizer of the convex ISO objective \eqref{eq:Estimator} is achieved at $(J_i, H_i) = (J^*_i, H^*_i)$, meaning that the true interactions present in the model are fully ``screened''. Indeed, the ISO is an empirical average of the inverse of the factors in the Gibbs measure; if $\mathcal{F}_i(J_i, H_i) = \exp(\sum_{j\neq i} J_{ij} \sigma_{i} \sigma_{j} + H_{i}\sigma_{i})$, then $S(J_i,H_i) = \langle \mathcal{F}_i^{-1}(J_i,H_i) \rangle_{M}$. In the limit of large number of samples $S(J_i,H_i) \rightarrow S^*(J_i,H_i) = \langle 1/\mathcal{F}_i(J_i,H_i) \rangle$. The derivative of the ISO corresponds to weighted pairwise correlations, $\partial S^{*} / \partial J_{ij} = \langle\sigma_i \sigma_j / \mathcal{F}_i(J_i,H_i)\rangle$, and this sheds light on its key property. When $(J_i, H_i) = (J^*_i, H^*_i)$, $\partial S^{*} / \partial J_{ij} \vert_{J^*_i,H^*_i} = 0$, meaning that the minimum of ISO is achieved at $(J_i, H_i) = (J^*_i, H^*_i)$ as $M \to \infty$.

\subsection*{High-temperature expansion of exact estimators and connection to mean-field}

Among all heuristics undertaking to solve the inverse Ising problem, a large fraction of methods is based on mean-field approximations using various level of sophistication, see \cite{Roudi2009} for a review. In particular, the first such attempt to solve the inverse Ising problem is based on a na\"{i}ve mean-field approach where inferred couplings are related to the inverse reduced correlation matrix \cite{kappen1998boltzmann}. Although these techniques provide satisfactory estimates in the high temperature regime, they are known to exhibit poor behaviors at low temperature when the model develops long-range correlations, even for an unlimited number of samples \cite{montanari2009graphical}.

It is interesting to observe that there exists a connection between mean-field approaches and the high-temperature expansion of the exact estimators RISE and RPLE. A second-order Taylor expansion of the Pseudo-Likelihood objective function (without regularizer) and the ISO around the high temperature point $(J_i,H_i)=(0,0)$ produces an explicitly solvable minimization problem, see Supplementary Text for an exact derivation. It is remarkable that in this regime both objective functions produce identical estimates for the model parameters. Couplings and magnetic fields reconstructed in this Mean-Field Regime (MFR) are expressed as functions of the inverse connected correlations matrix and local magnetizations
\begin{equation}
\label{eq:mfr_relations}
\widehat{J}_{ik}^{\text{MFR}} = -\frac{\left[\bar{C}^{-1}\right]_{ik}}{\left[\bar{C}^{-1}\right]_{ii}}, \quad \widehat{H}_{i}^{\text{MFR}} = -m_{i} + \sum_{j\neq i} \frac{\left[\bar{C}^{-1}\right]_{jk}}{\left[\bar{C}^{-1}\right]_{ii}} m_{j},
\end{equation}
where the matrix of empirical connected correlations and local magnetizations are direclty computed from samples using the formulae
 $\bar{C}_{ij} = \langle\sigma_{i}\sigma_{j}\rangle_M - \langle\sigma_{i}\rangle_M \langle\sigma_{j}\rangle_M$ and $m_{i} = \langle\sigma_{i}\rangle_M$.
Note that there is a subtle difference between the MFR estimates in Eq.~\eqref{eq:mfr_relations} and the na\"{i}ve mean-field estimates in \cite{kappen1998boltzmann}. The values produced by the na\"{i}ve mean-field method are directly equal to the inverse connected correlation matrix, whereas the MFR estimates are rescaled by the diagonal entries of this matrix. As a result, even though both estimators provide similar answers, the choice of symmetrization and thresholding procedures for the reconstructed parameters can lead to significant discrepancies in the final estimates of the graph structure. It is worth noticing that the exact same expression producing the MFR estimates arises in the context of reconstructing multivariate Gaussian distributions \cite{meinshausen2006high}. This parallel suggests that an optimal thresholding and symmetrization procedure for the MFR estimates is likely to be based on the geometrical mean rather than the arithmetic average.
\cite{misra2017towards}.

\begin{acknowledgments}
The authors are grateful to G. Bresler, C. Coffrin, A. Montanari, N. Uvarov and M. Zamparo for fruitful discussions and valuable comments. The work at LANL was carried out under the auspices of the National Nuclear Security Administration of the U.S. Department of Energy under Contract No. DE-AC52-06NA25396. The code implementing all exact estimators presented in this paper, as well as real data produced by the D-Wave quantum computer ``Ising'' at LANL and used in this work can be found at \url{https://github.com/lanl-ansi/inverse_ising}.
\end{acknowledgments}

\appendix*

\onecolumngrid

\renewcommand{\theequation}{S\arabic{equation}}

\setcounter{equation}{0}

\renewcommand{\thefigure}{S\arabic{figure}}

\setcounter{figure}{0}

\renewcommand{\thesection}{S\arabic{section}}

\setcounter{section}{0}

\renewcommand\appendixname{}

\vspace{1cm}

\begin{center}
\noindent {\bf {\Large Supplementary Text}}\\  
\end{center}

In the Supplementary Text, we cover different technical aspects of our methods both on the theoretical and practical side. First, we provide the mathematical analysis and proofs of exactness of the RPLE, RISE and logRISE algorithms in section \ref{sec:theory_estimators}. In section \ref{sec:optimization} we discuss implementation questions related to the minimization of the estimators. A comment on our procedure for $M^{*}$ selection is given in section \ref{sec:M_selection}. The implications of the mathematical analysis on the selection of the regularization parameter $\lambda$ is presented in section \ref{sec:theory_lambda}. This is followed by a description of our empirical procedure for selecting the hyperparameter $\lambda$ (section \ref{sec:lambda_selection}), as well as by a discussion on a possible cross-validation method for selecting $\lambda$ in practical setting when a sufficient number of samples is available (section \ref{sec:cross_validation}). In section \ref{sec:rise_scaling} the experimental results and scaling for the RISE are presented. Finally, details on derivation of the expansion of the estimators in the mean-field regime can be found in section \ref{sec:ht_expansion}.

\section{Analysis of the estimators RPLE, RISE and logRISE}
\label{sec:theory_estimators}

In this section we present a rigorous study of the trade-off between sample complexity and accuracy for both the RPLE and logRISE, and highlight the differences with the properties of the RISE. We start our analysis with RPLE and RISE which belong to the class of the so-called M-estimators i.e. estimators resulting from the minimization of an empirical average of convex functions \cite{negahban2009unified}. The mathematical framework that we use combines the techniques from the theory of M-estimators and the analysis of the RISE, developed in \cite{screening2016theoretical}. Here, we apply the key points of this theory to the analysis of the RPLE in order to provide a better understanding of the performance discrepancy between the two estimators.

For convenience, let us bring the form of both estimators to uniformity. Maximizing the local pseudo-likelihood objective function associated with node $i$ is equivalent to minimizing its opposite:  
\begin{equation}
\widetilde{\mathcal{L}}_{i}(J_{i},H_{i})=-\mathcal{L}_{i}(J_{i},H_{i})=\langle\ln(1+\exp(-2\sigma_{i}(H_{i}+\sum_{j\neq i}J_{ij}\sigma_{j}))\rangle_{M},
\label{eq:PL_objective}
\end{equation}
where the empirical average is defined as
\begin{equation}
\langle f(\sigma)\rangle_{M}=\frac{1}{M}\sum_{m=1}^{M}f(\sigma^{(m)}),
\end{equation}
and $J_{i}$ is the shortcut notation for $\{J_{ij}\}_{j \neq i}$. The outcome of the RPLE is simply the minimizer of the negative regularized pseudo-likelihood objective function
\begin{equation}
(\widehat{J}_{i}^{\text{RPLE}},\widehat{H}_{i}^{\text{RPLE}})=\argmin_{(J_{i},H_{i})}\left[\widetilde{\mathcal{L}}_{i}(J_{i},H_{i})+\lambda\Vert J_{i}\Vert_{1}\right].
\label{eq:RPLE}
\end{equation}
The RISE is based on the \emph{Interaction Screening Objective} (ISO)
\begin{equation}
S_{i}(\widehat{J}_{i},\widehat{H}_{i})=\langle\exp(-\sum_{j\neq i}J_{ij}\sigma_{i}\sigma_{j}-H_{i}\sigma_{i})\rangle_{M}.\label{eq:sup_mat_ISO}
\end{equation}
The original form of RISE introduced in \cite{screening2016theoretical} reads 
\begin{equation}
(\widehat{J}_{i}^{\text{RISE}},\widehat{H}_{i}^{\text{RISE}})=\argmin_{(J_{i},H_{i})}\left[S_{i}(J_{i},H_{i})+\lambda\Vert J_{i}\Vert_{1}\right].\label{eq:sup_mat_RISE}
\end{equation}
Notice that in the main text we have mainly studied the logRISE, for which $S_{i}$ is replaced by $\ln S_{i}$. While the original form of the estimator is more amenable to the theoretical analysis, the logarithmic version is more suitable for the implementation due to its numerical stability, and even requires less samples in hard cases. A detailed empirical comparison between these two versions of the RISE is provided later.

For the sake of simplicity and the clarity of presentation, in our analysis we consider the case where magnetic fields are set to zero. Our main result of the error analysis of the RPLE is contained in the following theorem. For completeness, we also present the corresponding result for the RISE.
\begin{thm*}
Let $M$ be the number of i.i.d. samples from an Ising model with
$N$ variables, bounded degree $d$ and maximum coupling $\beta=\max_{ij}\left|J_{ij}^{^{*}}\right|$.
The reconstruction error on the couplings (in the neighborhood of node $i$) of the RPLE with regularization parameter $\lambda=c_{1}\sqrt{\ln\left(N^{2}/\epsilon\right)/M}$ is bounded with probability $1-\epsilon/N$ as
\[
\left\Vert \widehat{J}_{i}^{\text{RPLE}}-J^{*}_{i}\right\Vert _{2}\leq C_{d}e^{4\beta d}\sqrt{\frac{\ln\left(N^{2}/\epsilon\right)}{M}}.
\]
For the RISE, the same error is estimated as
\[
\left\Vert \widehat{J}_{i}^{\text{RISE}}-J^{*}_{i}\right\Vert _{2}\leq C'_{d}e^{3\beta d}\sqrt{\frac{\ln\left(N^{2}/\epsilon\right)}{M}},
\]
where $C_{d}$ and $C'_{d}$ depend only polynomially on $d$, and $c_{1}$ is a constant.
\end{thm*}

The control of the error on the reconstructed couplings is important for the following reason: if this error is smaller than (say) $\alpha/2$, where $\alpha=\min_{ij \in E}\left|J_{ij}^{^{*}}\right|$, it becomes easy to reconstruct the structure of the neighborhood of node $i$ by declaring the edges whose reconstructed coupling is less than $\alpha/2$ in absolute value, to be absent. Repeating this procedure over $N$ neighborhoods, we can guarantee (through the union bound) the exact reconstruction of the graph with probability $1-\epsilon$ (that is the reason why the level of error in the Theorem is required with a smaller probability $1-\epsilon/N$ for each neighborhood). Given the graph structure, it is then easier to estimate the values of the non-zero couplings. The results of the Theorem above allow to estimate the number of samples $M^{*}$ which is sufficient to obtain a fixed error on the couplings, and hence to recover the structure of the graph, for both estimators: $M^{*}\propto e^{8\beta d}\ln N$ for the RPLE and $M^{*}\propto e^{6\beta d}\ln N$ for the RISE. Below we sketch the proof of the Theorem, and highlight the differences\ in the nature of the estimators which explains their distinct performance in practice. As argued in the main text, the expressions for the errors given above are not tight, and represent the upper bounds on the actual required number of samples; the detailed numerical experiments presented in this paper show that the scalings of $M$ in practice is better than the theoretically predicted ones for both estimators.

\begin{figure}[!tb]
\begin{centering}
\includegraphics[width=0.5\columnwidth]{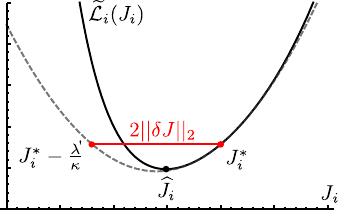}
\par\end{centering}
\caption{The objective function, $\widetilde{\mathcal{L}}_{i}\left(J_{i}\right)$,
is shown in black. The quadratic lower-bound $Q_{\widetilde{\mathcal{L}}_{i}}$ centered in $J^{*}_{i}$ is the
gray dashed line. The estimated distance between $J^{*}_{i}$ and $\widehat{J}$
is indicated by the red line starting in $J^{*}_{i}$. As $\widetilde{\mathcal{L}}_{i}\left(J_{i}\right)$ is convex, the minimum point is ensured to
be enclosed between the quadratic lower-bound and the red line, which gives a way to estimating the difference $\Vert \widehat{J}_{i} - J^{*}_{i} \Vert_{2}$, as explained in the text. The similar proof applies to the case of the Interaction Screening Objective $S_{i}\left(J_{i}\right)$ \label{fig:proof_strategy_m_estimator} }
\end{figure}

\subsection{Analysis of the RPLE}

To bound the distance between the true parameters of the model $J^{*}_{i}$ and their estimated counterparts $\widehat{J}_{i}$ for finite $M$, we use a proof strategy based on constructing a quadratic lower-bound of the objective function centered around $J^{*}_{i}$. In the case of the objective function of the RPLE type, an explicit form of the quadratic lower-bound $Q_{\widetilde{\mathcal{L}}_{i}}$ which satisfies $\widetilde{\mathcal{L}}_{i}\left(J_{i}\right)
\geq Q_{\widetilde{\mathcal{L}}_{i}}\left(J_{i}\right)$ can be evaluated, see \cite{screening2016theoretical} for the detailed description of the procedure. The idea is that the distance $\Vert \delta J \Vert_{2} \equiv \Vert \widehat{J}_{i} - J^{*}_{i} \Vert_{2}$ can be estimated using this explicit form of $Q_{\widetilde{\mathcal{L}}_{i}}$ and the fact that the estimator is convex. This quadratic lower-bound  is approximately given by a second-order Taylor expansion of $\widetilde{\mathcal{L}}_{i}(\widehat{J}_{i})=\widetilde{\mathcal{L}}_{i}(J^{*}_{i}+\delta J)$ around $J^{*}_{i}$:
\begin{equation}
Q_{\widetilde{\mathcal{L}}_{i}}\left(J^{*}_{i}+\delta J\right) \approx \widetilde{\mathcal{L}}_{i}\left(J^{*}_{i}\right)+\left\langle \delta J,\nabla\widetilde{\mathcal{L}}_{i}\left(J^{*}_{i}\right)\right\rangle +\frac{1}{2}\left\langle \delta J,\nabla^{2}\widetilde{\mathcal{L}}_{i}\left(J^{*}_{i}\right)\delta J\right\rangle .
\label{eq:quadratic_expansion}
\end{equation}

Since $\widehat{J}_{i}$ realizes the minimum of the estimator $\widetilde{\mathcal{L}}_{i}(J_{i})$, we have $\widetilde{\mathcal{L}}_{i}(J^{*}_{i}) \geq \widetilde{\mathcal{L}}_{i}(\widehat{J}_{i})$ (where the equality occurs for $M \to \infty$, when $\widehat{J}_{i}$ coincides with $J^{*}_{i}$). Because $\widetilde{\mathcal{L}}_{i}  \geq Q_{\widetilde{\mathcal{L}}_{i}}$, the convex sublevel set of $\widetilde{\mathcal{L}}_{i}$ corresponding to the value $\widetilde{\mathcal{L}}_{i}(J^{*}_{i})$ is contained in the convex sublevel set of $Q_{\widetilde{\mathcal{L}}_{i}}$, and the minima $\widehat{J}_{i}$ must lie within this region:
\begin{align} \label{eq:level_sets}
    \widehat{J}_{i} \in \{J_i \vert \widetilde{\mathcal{L}}_{i}(J_i) \leq  \widetilde{\mathcal{L}}_{i}(J_i^*) \} \subseteq \{J_i \vert Q_{\widetilde{\mathcal{L}}_{i}}(J_i) \leq  Q_{\widetilde{\mathcal{L}}_{i}}(J_i^*) \}.
\end{align}
As a result, the distance 
$\Vert \widehat{J}_{i} - J^{*}_{i} \Vert_{2}$ can be upper bounded by the diameter of the convex region on the right hand side of \eqref{eq:level_sets}. 
This idea is sketched in the Fig.~\ref{fig:proof_strategy_m_estimator} as a one-dimensional representation. Here, the quadratic expansion \eqref{eq:quadratic_expansion} reads: $Q_{\widetilde{\mathcal{L}}_{i}}\left(J^{*}_{i}+\delta J\right)
\approx\widetilde{\mathcal{L}}_{i}\left(J^{*}_{i}\right) + \lambda' \delta J +\frac{1}{2} \kappa (\delta J)^{2}$. This function takes the value $\widetilde{\mathcal{L}}_{i}\left(J^{*}_{i}\right)$ at two points: $\delta J = 0$ and $\delta J = - 2 \lambda' / \kappa$. The distance between the estimated and the true parameters can be hence estimated as
\begin{equation}
\Vert \widehat{J}_{i} - J^{*}_{i} \Vert_{2} \leq \frac{\lambda'}{\kappa}.
\label{eq:error_couplings}
\end{equation}
In the high-dimensional setting, $\lambda'$ represents the largest component of the gradient $\nabla\mathcal{L}$, and $\kappa$ is the smallest eigenvalue of the Hessian matrix $\nabla^{2}\mathcal{L}$, both evaluated at the point $J^{*}$. Given this proof strategy, we need to estimate $\lambda'$ and $\kappa$ 
in order to recover the precise statement of the Theorem.

{\bf Estimation of $\lambda'$:} It is straightforward to compute the gradient of the pseudo-likelihood objective \eqref{eq:PL_objective}: 
\begin{equation}
\nabla\widetilde{\mathcal{L}}_{i}\left(J^{*}_{i}\right)=\langle\sigma_{\backslash i}(\tanh(\sum_{j\in\partial i}J_{ij}^{*}\sigma_{j})-\sigma_{i})\rangle_{M},
\label{eq:gradient_RPLE}
\end{equation}
where $\sigma_{\backslash i}$ denotes the vector of size $p-1$ containing
all spins but $\sigma_{i}$, and $\partial i$ denotes the set of neighbors of node $i$. As all components of the gradient at $J^{*}_{i}$ are
upper-bounded
\begin{equation}
\left|\frac{\partial}{\partial J_{ik}}\widetilde{\mathcal{L}}_{i}\left(J^{*}_{i}\right)\right|\leq2,
\end{equation}
we use Hoeffding's concentration inequality \cite{hoeffding1963probability} to show that any given gradient component is bounded with high probability
\begin{equation}
\mathbb{P}\left[\left|\frac{\partial}{\partial J_{ik}}\widetilde{\mathcal{L}}_{i}\left(J^{*}\right)\right|\geq\frac{4t}{\sqrt{M}}\right]\leq c_{1}e^{-t^{2}},
\label{eq:hoeffding}
\end{equation}
where $c_{1}>0$ is a constant. The inequality \eqref{eq:hoeffding} means that the gradient components lie in an interval with size of order $4/\sqrt{M}$. Moreover, the probability that these gradient components lie outside of this interval, and are away by a multiplicative factor $t$ decreases exponentially in $t^2$. By choosing $t=\sqrt{\ln\left(N^{2}/\epsilon\right)}$, we limit the right hand side in \eqref{eq:hoeffding} by $\epsilon/N^{2}$. This shows that with probability at least $1-\epsilon/N^{2}$ any given gradient components is upper-bounded
\begin{equation}
\left|\frac{\partial}{\partial J_{ik}}\widetilde{\mathcal{L}}_{i}\left(J^{*}\right)\right|\leq c_{1}\sqrt{\frac{\ln\left(N^{2}/\epsilon\right)}{M}}.
\label{eq:sup_mat_gradient}
\end{equation}
Recall that there are $N-1$ components of the gradient vector; taking the union bound over them, we can guarantee that the maximum over these $N-1$ gradient components is of the same order as in \eqref{eq:sup_mat_gradient} with probability $1-\epsilon/N$. Therefore, the quantity $\lambda'$ defined above can be estimated as
\begin{equation}
\lambda' \propto \sqrt{\frac{\ln\left(N^{2}/\epsilon\right)}{M}}.
\label{eq:lambda_prime}
\end{equation}

{\bf Estimation of $\kappa$:} The Hessian matrix of the pseudo-likelihood function $\widetilde{\mathcal{L}}$ can be found by direct computation and reads
\begin{equation}
\nabla^{2}\widetilde{\mathcal{L}}_{i}\left(J^{*}\right)=\langle\sigma_{\backslash i}\sigma_{\backslash i}^{\top}(1-\tanh(\sum_{j\in\partial i}J_{ij}^{*}\sigma_{j})^{2}\rangle_{M}.\label{eq:sup_mat_hessian_PL}
\end{equation}
Using the inequality $1-\tanh\left(x\right)^{2}\geq\exp\left(-2\left|x\right|\right)$
and the fact that $\left|\sum_{j\in\partial i}J_{ij}^{*}\sigma_{j}\right|\leq\beta d$, we show that the Hessian is lower-bounded in the positive semi-definite sense
\begin{equation}
\nabla^{2}\widetilde{\mathcal{L}}_{i}\left(J^{*}\right)\succeq e^{-2\beta d}C_{M},\label{eq:sup_mat_hessian_PL_bound}
\end{equation}
where the matrix $C_{M}$ is the empirical covariance matrix
\begin{equation}
C_{M}=\langle\sigma_{\backslash i}\sigma_{\backslash i}^{\top}\rangle_{M}.\label{eq:sup_mat_emp_cov}
\end{equation}
In expectation $C_{M}$ is equal to the covariance matrix $C=\langle\sigma_{\backslash i}\sigma_{\backslash i}^{\top}\rangle$ for which all eigenvalues are bigger than $a_C e^{-2\beta d}$ \cite{montanari2015computational}, where $a_C$ is a constant depending polynomially on $d$. However, already from the expression \eqref{eq:lambda_prime} we see that $M$ scales as $\ln N$ in order to guarantee the constant error on the couplings. In this so-called high-dimensional regime $M\propto \ln N$, the empirical covariance matrix possesses only $\mathcal{O}(\ln N)$ non-zero eigenvalues. The reason for $C_{M}$ to be severely rank deficient is that $C_{M}$ is the sum of $M$ rank-one matrices $\sigma_{\backslash i}^{(m)}\sigma_{\backslash i}^{\left(m\right)\top}$. Therefore the rank of $C_{M}$ can not exceed $M$.

This problem is circumvented by the presence of the $\ell_{1}$ penalty term in the optimization formulation of the RPLE \eqref{eq:RPLE}. It turns out that if the penalty parameter $\lambda$ is greater than the largest component of the gradient \eqref{eq:lambda_prime} (which explains why we denoted the bound on the gradient components as $\lambda'$), the only relevant eigenvalues correspond to eigenvectors that are sparse; see \cite{negahban2009unified} for a more precise statement. Such eigenvalues are called restricted eigenvalues as they correspond to the minimum of the quadratic form associated with $C_M$ restricted to the sector of sparse vectors. An intuitive explanation of this property is that perturbations of $J^{*}_{i}$ with $\delta J$ which are not sparse drastically change the value of the $\ell_{1}$ penalty. Therefore a non-sparse perturbation $\delta J$ increases the value of the pseudo-likelihood objective $\widetilde{\mathcal{L}}$ with $\ell_{1}$ penalty even though it may not change the value of the pseudo-likelihood objective alone, which discourages such directions of perturbation.

It remains to verify that restricted eigenvalues of $C_{M}$ are with high probability bounded-away from zero. A technical proof of this statement can be found in \cite{screening2016theoretical} where it is shown that with high probability $C_{M}$ has all its restricted eigenvalues greater than $\frac{1}{2} a_C e^{-2\beta d}$. Combining this bound with \eqref{eq:sup_mat_hessian_PL_bound}, we get the following estimation of $\kappa$:
\begin{equation}
\kappa = \frac{1}{2} a_C e^{-4\beta d}.
\label{eq:kappa}
\end{equation}
Now using the expression \eqref{eq:error_couplings}, we finally determine the error between the couplings and their estimated counterpart as the ratio between \eqref{eq:lambda_prime} and \eqref{eq:kappa}: 
\begin{equation}
\left\Vert \widehat{J}_{i}^{\text{RPLE}}-J_{i}\right\Vert _{2} \leq \frac{\lambda'}{\kappa} \propto e^{4\beta d}\sqrt{\frac{\ln\left(N^{2}/\epsilon\right)}{M}}.\label{eq:sup_mat_PLE_error}
\end{equation}
This final inequality represents the first statement of the Theorem. It also shows that a constant error on the couplings, and hence the structure recovery, can be obtained with $M^{*}\propto e^{8\beta d}\ln N$.

\subsection{Analysis of the RISE}

We now proceed with a similar analysis on the RISE \eqref{eq:sup_mat_RISE}. The gradient of the Interaction Screening Objective reads
\begin{equation}
\nabla S_{i}\left(J^{*}_{i}\right)= - \langle\sigma_{\backslash i}\exp(-\sum_{j\in\partial i}J_{ij}^{*}\sigma_{i}\sigma_{j})\rangle_{M}.\label{eq:sup_mat_gradient_ISO}
\end{equation}
Unlike for the pseudo-likelihood objective, components of the gradient of the ISO are not bounded by a constant, but depend on $\beta$
\begin{equation}
\left|\frac{\partial}{\partial J_{ik}}S_{i}\left(J^{*}\right)\right|\leq e^{\beta d}.\label{eq:sup_mat_support_grad_ISO}
\end{equation}
Here a direct application of Hoeffding's concentration inequality
would produce a bound on the gradient that scales with $e^{\beta d}$.
It would further imply that the $\ell_{1}$-penalty parameter $\lambda$
has to scale with $e^{\beta d}$, which is not a desirable property
for practical implementations as $\beta$ and $d$ are often unknown.
Fortunately, it is possible to obtain a tighter estimate by taking into account
the variance of the ISO in Eq. (\ref{eq:sup_mat_gradient_ISO}). We observe that the variance of any component is constant and
equal to one
\begin{align}
\mathrm{Var}[\nabla S_{i}\left(J^{*}\right)] & =\langle\exp(-2\sum_{j\in\partial i}J_{ij}^{*}\sigma_{i}\sigma_{j})\rangle=1,\label{eq:sup_mat_var_iso_polyakov}
\end{align}
where in the last step we perform the change of variable $\sigma_{i}\rightarrow-\sigma_{i}$ while computing the expectation. This remarkable property of Ising models contained in Eq. (\ref{eq:sup_mat_var_iso_polyakov})
has already been noticed a long time ago by Polyakov in the context of “disorder” parameter analysis of the Ising models, see e.g. last chapter of \cite{polyakov1987gauge}. Using Bernstein's concentration inequality we take advantage of the fact that the gradient of the ISO has a variance  (\ref{eq:sup_mat_var_iso_polyakov}) much smaller than its support  (\ref{eq:sup_mat_support_grad_ISO}). Hence if the number of samples is at least of order $M\propto e^{2\beta d}\ln N$, the gradient of the ISO concentrates as fast as the gradient of $\widetilde{\mathcal{L}}$ in \eqref{eq:sup_mat_gradient}.

The notable difference between the RISE and the RPLE comes from the analysis of the Hessians of their objective functions. After a straightforward computation we find that the Hessian of the ISO reads
\begin{equation}
\nabla^{2}S_{i}\left(J^{*}_{i}\right)=\langle\sigma_{\backslash i}\sigma_{\backslash i}^{\top}\exp(-\sum_{j\in\partial i}J_{ij}^{*}\sigma_{i}\sigma_{j})\rangle_{M}.\label{eq:sup_mat_hessian_iso}
\end{equation}
As $\left|\sum_{j\in\partial i}J_{ij}^{*}\sigma_{i}\sigma_{j}\right|\leq\beta d$,
the Hessian of the ISO is lower-bounded in the positive semi-definite
sense by the empirical covariance matrix
\begin{equation}
\nabla^{2}S_{i}\left(J^{*}_{i}\right)\succeq e^{-\beta d}C_{M}.\label{eq:sup_mat_hessian_iso_bound}
\end{equation}
Therefore, the formula \eqref{eq:error_couplings} gives us the guarantee with probability $1-\epsilon/N$ that couplings are estimated within the following error
\begin{equation}
\left\Vert \widehat{J}_{i}^{\text{RISE}}-J_{i}\right\Vert _{2} \leq \frac{\lambda'}{\kappa} \propto e^{3\beta d}\sqrt{\frac{\ln\left(N^{2}/\epsilon\right)}{M}}.\label{eq:sup_mat_RISE_error}
\end{equation}
This implies that the RISE recovers couplings up to a given constant accuracy
with a number of samples that scales as $M^{*}\propto e^{6\beta d}\ln N$.
This scaling is $e^{2\beta d}$ faster compared to the one found for the RPLE.

\subsection{Analysis of the logRISE}

The estimator logRISE is not a M-estimator, i.e. a minimizer of an averaged function, but the logarithm of such estimator
\begin{equation}
(\widehat{J}_{i}^{\text{RISE}},\widehat{H}_{i}^{\text{RISE}})=\argmin_{(J_{i},H_{i})}\left[\ln(S_{i}(J_{i},H_{i}))+\lambda\Vert J_{i}\Vert_{1}\right],\label{eq:sup_mat_logRISE}
\end{equation}
where $S_{i}(J_{i},H_{i})$ is the Interaction Screening Objective from Eq~\eqref{eq:sup_mat_ISO}.
For this reason the method employed to analyze RISE and RPLE can be essentially carried away for logRISE with the drawback that the bounds derived within this framework are potentially much more loose.
We reproduce here the key points of the analysis which are upper-bounding the gradient and lower-bounding the hessian of the objective function.

The gradient of logRISE at the point $J_{i} = J^{*}_{i}$ reads as follow
\begin{equation}
\nabla \ln(S_{i}\left(J^{*}_{i}\right))= (S_{i}\left(J^{*}_{i}\right))^{-1}\nabla S_{i}\left(J^{*}_{i}\right).\label{eq:sup_mat_gradient_logRISE}
\end{equation}
As the ISO is bounded away from below and from above at $J^{*}_{i}$
\begin{equation}
 e^{- \beta d} \leq S_{i}\left(J^{*}_{i}\right) \leq e^{\beta d},
\end{equation}
it implies that the gradient components of logRISE can be at most $e^{\beta d}$ larger than the corresponding gradient components of RISE. We highlight that this crude estimate neglects the strong correlations existing between the function $S_{i}\left(J^{*}_{i}\right)$ and its gradient $\nabla S_{i}\left(J^{*}_{i}\right)$. For instance note that the ISO is always larger than any gradient components $S_{i}\left(J^{*}_{i}\right) \geq \Vert \nabla S_{i}\left(J^{*}_{i}\right)\Vert_{\infty}$. Notice that in practice these correlations may facilitate reconstruction in certain models. In a similar fashion one can also show that the expected hessian of logRISE has bounded eigenvalues
\begin{equation}
\mathbb{E}\left[\nabla^{2}\ln(S_{i}\left(J^{*}_{i}\right))\right] \succeq e^{-2\beta d}C_{M},
\end{equation}
which are smaller than for RISE by a factor at most $e^{-\beta d}$.

These bounds give us the guarantee with probability $1-\epsilon/N$ that couplings are estimated within an error
\begin{equation}
\left\Vert \widehat{J}_{i}^{\text{logRISE}}-J_{i}\right\Vert _{2} \propto e^{5\beta d}\sqrt{\frac{\ln\left(N^{2}/\epsilon\right)}{M}}.\label{eq:sup_mat_logRISE_error}
\end{equation}
Therefore the number of samples requires by logRISE for perfect structure reconstruction scales with the inverse temperature at most like $e^{10\beta d}$.

\section{On optimization techniques for minimizing the estimators}
\label{sec:optimization}

In the algorithmic implementation, it might be convenient to pass the $\ell_{1}$ regularization as a constraint to the optimization problem in the slack form: for example, the expression \eqref{eq:sup_mat_RISE} can be rewritten as
\begin{equation}
(\widehat{J}_{i},\widehat{H}_{i}) = \arg \min_{(J_i,H_i)} \Big[  S_{i}(J_i,H_i) + \lambda \sum_{j=1}^{N} \rho_{j} \Big]
\end{equation}
with the constraints
\begin{equation}
J_{ik} \leq \rho_{k}, \quad J_{ik} \geq -\rho_{k} \quad \text{for} \quad k \neq N.
\end{equation}
In both cases, the algorithms can be initialized with the values of all the parameters equal to zero, $J_{ij}=0$ for all $(ij)$ and $H_{i}=0$ for all $i \in V$, which corresponds to the value $S_{i}(\underline{0}_{i},0)=1$ for all $i \in V$. Notice that it is possible to additionally impose the constraint $S_{i}(J_{i},H_{i}) \leq 1$ or $\log S_{i}(J_{i},H_{i}) \leq 0$ for ensuring the numerical stability of the algorithms RISE and logRISE, respectively.

The numerical results presented in this work have been obtained using the Ipopt solver \cite{biegler2009large}. Our additional tests (not shown) indicate that for the large-scale problems, the use of the first-order composite gradient descent methods \cite{nesterov2007gradient, agarwal2010fast} is preferable, since it achieves the computational complexity $\mathcal{O}(M N^{2})$ compared to the complexity $\mathcal{O}(M N^{4})$ of the general convex solvers that use matrix inversion as a subroutine. 
The basic building block of the first order methods is the so-called proximal gradient method that is used for minimizing non-differentiable convex functions. This method has gained much popularity for optimizing $\ell_1$ regularized convex objectives \cite{beck2009fast}, and both logRISE and RPLE fall in this category. The resulting algorithm takes the form of a gradient descent step of the unregularized objective with an appropriately chosen step size, followed by a soft-thresholding step to account for the non-differentiable $\ell_1$ term.
Other possible implementation improvements of the reconstruction algorithms include the use of the stochastic gradient descent and parallelization (since the problem is solved independently for each node).

\section{On the procedure for $M^{*}$ selection}
\label{sec:M_selection}

In this section, we show that repeated successful runs over $L=45$ different sets of samples is required for in the numerical experiments in order to guarantee that the graph is reconstructed with probability above $1 - \epsilon = 0.95$ for our choice $\epsilon = 0.05$ with confidence at least $90\%$. Indeed, in our case the numerical experiment is equivalent to generating flips of an unfair coin with probability of success equal to $p$. Assuming the uniform initial prior, let us denote by $P_{\text{posterior}}(p \mid L)$ the posterior probability over $p$ after a series of $L$ successful reconstructions, which is given by the Beta distribution for this Bernoulli process. Let us define
\begin{equation}
p_{\text{conf}} \equiv \int_{1-\epsilon}^{1}P_{\text{posterior}}(p \mid L_{\text{succ}} = L) dp.
\label{confidence}
\end{equation}
We require that $p_{\text{conf}} > 0.9$, and use Eq.~\eqref{confidence} for determining the necessary $L$. It is easy to check that for $L=45$ we obtain $p_{\text{conf}} = 0.905532$. This value of $L$ has been used in the computations of all points in the scaling plots of the main text, as well as in Fig.~S2 below.

\section{On the theoretical predictions for $\lambda$ selection}
\label{sec:theory_lambda}

Our analysis of RISE, logRISE and RPLE provides certain guarantees for the value of the regularizer parameter $\lambda$. Any regularizer parameter larger than the objective gradient fluctuations is promoting sparsity. Even though this guarantee is certainly conservative and a smaller $\lambda$ may be much more efficient in practice, it enables us to make interesting predictions. The value of $\lambda$ scales with the number of samples and the size of the system at most like $\lambda \propto \sqrt{\frac{\ln N}{M}}$. For the RISE, the regularizer $\lambda$ does not scale with the degree or the coupling strength of the Ising model as the variance of the ISO is independent of these parameters. For RPLE, the story is a little different. In general, fluctuations of the PL objective gradient Eq.~\eqref{eq:gradient_RPLE} do not grow with the couplings strength or the maximum degree. This ensure that a $\lambda$ independent of these quantities is promoting sparsity in the RPLE. However on close inspection, it is possible that for specific Ising models, such as ferromagnetic systems, fluctuations of the gradient are decreasing exponentially fast with $\gamma$. This suggest that the optimal regularizer parameter for RPLE depends on the couplings strength and nodal degrees in a non-trivial manner and becomes smaller as these two quantities grow. Unfortunately, finding the \emph{a priori} optimal $\lambda$ for RPLE can be difficult in practice as the system (and thus $\gamma$) is not known in advance. Finally note that the loose bound on the gradient fluctuations for logRISE grows exponentially fast with $\gamma$. Nevertheless our numerical studies show that choosing a regularizer $\lambda$ independent of $\gamma$ is sufficient for promoting sparsity with logRISE.

\begin{figure}[!htb]
\begin{center}
\includegraphics[width=\columnwidth]{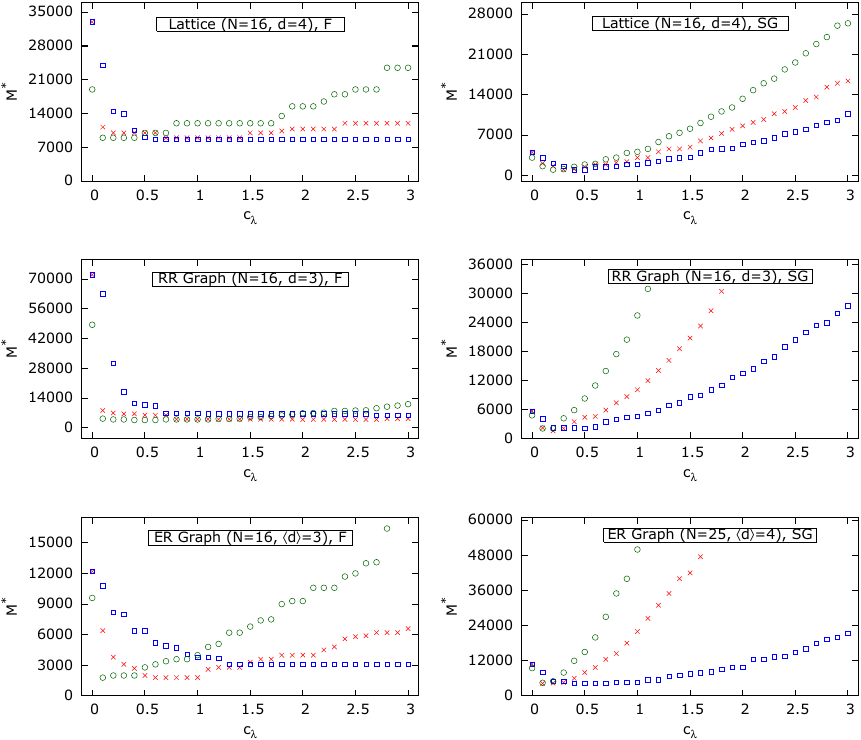}
\caption{The required number of samples $M^{*}$ for the RPLE (green circles), the RISE (red crosses) and the logRISE (blue squares) as a function of $c_{\lambda}$ on different topologies: square lattice, random regular graph with $d=3$, Erd{\"o}s-R{\'e}nyi graphs with $\langle d\rangle = 3$ and $\langle d\rangle = 4$ (from top to bottom). In these plots, the original couplings $J^*$ have been randomly generated assuming ferromagnetic (denoted ``F'', left column) and spin glass (denoted ``	SG'', right column) models without magnetic field taking absolute values $\vert J^{*}_{ij} \vert$ in the following ranges: $[0.3,0.7]$ for the square lattice, and $[0.4,1.0]$ for random regular and Erd{\"o}s-R{\'e}nyi graphs.}
\label{fig:lambda_selection}
\end{center}
\end{figure}

\section{Empirical selection of the regularization parameter $\lambda$}
\label{sec:lambda_selection}

In this section, we run extensive simulations on different graph topologies in order to determine the optimal consensus value of the regularization parameter $\lambda$ for the three estimators RPLE, RISE and logRISE.

As it follows from our theoretical analysis above, the correct scaling of $M^{*}$ with the model parameters is guaranteed if one takes $\lambda > \lambda'$, where $\lambda'$ is given by the expression \eqref{eq:lambda_prime}. Although giving a sufficient condition, the expression for $\lambda'$ is not guaranteed to be tight, as we discussed in the previous section, especially with respect to the constant coefficients. At the same time, the generic form in Eq. \eqref{eq:lambda_prime} is rather intuitive: as is usual for the law of large numbers, $\lambda'$ is inversely proportional to the square root of the number of samples $M$ which controls the concentration of the gradient of the objective function, and grows with $\ln N^{2}/\epsilon$, where $\epsilon$ is the required fixed error of reconstruction, and $N^{2}$ reflects the requirement of correctly estimating $N$ parameters of the $N$ nodes in the graph \cite{screening2016theoretical}. Hence, in what follows we study numerically the effect of application of the regularization term with the coefficient $\lambda$ in the form
\begin{equation}
\lambda = c_{\lambda}\sqrt{\frac{\ln(N^2/\epsilon)}{M}}
\label{lambda_numerical}
\end{equation}
for a range of $c_{\lambda}$. Our goal is to determine an appropriate consensus value of $c_{\lambda}$ for different ensembles of Ising models. Of course, in practice the training over $c_{\lambda}$ can be restricted to a specific family of Ising models to which the given model to be recovered is supposed to belong, if this prior information is available.  

The results are presented in the Fig.~\ref{fig:lambda_selection} for different topologies (grid with periodic boundary conditions, random regular graphs and Erd{\"o}s-R{\'e}nyi graphs) in ferromagnetic and spin glass regimes. First of all, we notice that the choice of the estimator in the form \eqref{eq:sup_mat_logRISE} leads to more stable and smooth reconstruction performance with respect to the variation of $\lambda$ compared to both RPLE \eqref{eq:RPLE} and the original form of RISE \eqref{eq:sup_mat_RISE}; this observation is especially striking in the case of the spin glass models on random graphs, where the range of optimal $c_{\lambda}$ for the estimators \eqref{eq:RPLE}-\eqref{eq:sup_mat_RISE} appears to be much narrower. Second, the behavior of $M^{*}$ as a function of $c_{\lambda}$ for the logRISE seems to be different in the cases of ferromagnetic and spin glass models: while in the case of interactions of ferromagnetic type the optimal values of the regularization coefficient $\lambda$ are achieved for larger $c_{\lambda}$, the spin glass model requires lower values of $c_{\lambda}$ for a correct topology recovery. Based on these conclusions, in all cases we have chosen to use the expression \eqref{lambda_numerical} with the consensus values of the coefficient $c_{\lambda}=0.8$ for logRISE and $c_{\lambda}=0.4$ for RISE which are not optimal for any given model, but yield a reasonably optimized performance for a wide range of different topologies and model types. The corresponding value used for the RPLE has been chosen as $c_{\lambda}=0.2$, which is close to the optimal value for this estimator in the majority of the cases tested. These values of regularizer have been used in the simulations throughout the paper.

\begin{figure}[!htb]
\begin{center}
\includegraphics[width=\columnwidth]{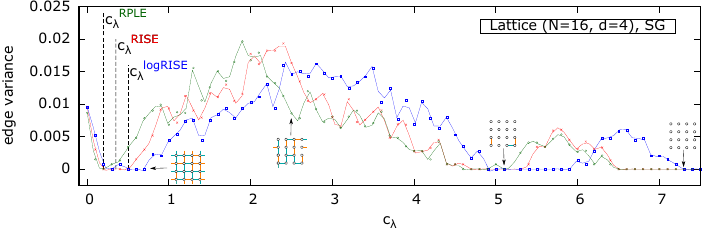}
\caption{Selection of the hyperparameter $\lambda$ through the $K$-fold cross-validation method on a $4\times4$ spin-glass system on a square lattice. Originally provided $11000$ samples have been split equally in $10$ folds. As $M^{*}\approx 1000$ for this system, each fold contains enough samples for a perfect reconstruction for some value of $\lambda$. The figure depicts the average score per edge with respect to $c_{\lambda}$ that is related to the hyperparameter through the formula $\lambda = c_{\lambda} \sqrt{\ln(N^2/\epsilon)/M}$. The value of $\epsilon$ is chosen to be $0.05$ as usual. The behavior of the reconstruction algorithms RPLE, RISE and logRISE are shown in green, red and blue respectively. For certain values of $c_{\lambda}$, we draw the graph spanned by edges that have been consistently reported as being present with a score of $0$. We can clearly identify three ``islands of consistency" on the $c_\lambda$ axis. We have also reported the optimal $c_\lambda$ found by exhaustive search for the three algorithms.}
\label{fig:lambda_cross_validation}
\end{center}
\end{figure}

\section{Hyperparameter $\lambda$ selection through cross-validation}
\label{sec:cross_validation}

In this section we consider selecting the $\ell_1$ regularizer parameter $\lambda$ through a cross-validation method. The procedure is simple and can be seen as a variant of a standard $K$-fold cross-validation. The technique consists in splitting the samples in $K$ smaller subsets of equal size, the folds. For a specific $\lambda$, we perform a graph reconstruction for each of the folds and then compare the consistency of the results. More precisely we compute the empirical probability $p$ of a candidate edge being identified as present over $K$ reconstructions. A score $p(1-p)$ is then assigned to the candidate edge that corresponds to the empirical variance of its reconstructions. If a candidate edge gets a score of $0$, it indicates that it has been consistently reconstructed as being present or absent over the $K$ reconstructions. If the score is higher, it means that the edge has been reconstructed differently over the $K$ reconstructions. The worst possible score of $0.25$ indicates that the edge has been uniformly at random identified as being present and absent. Our numerical results are reported in Figure~\ref{fig:lambda_cross_validation}. We can clearly identify regions in the $\lambda$ space for which the $K$ reconstructions have a $0$ score. We named these regions ``islands of consistency". When the hyperparameter is too large it is not surprising that a perfect $0$ score is achieved since the $\ell_{1}$ regularization is too strong and the algorithm reconstructs an empty graph consistently. On the other hand, we see that an absence of regularization ($\lambda=0$) results in a poor score as expected. A promising strategy consists in choosing the island of consistency associated with the samllest $\lambda$. We see that each algorithm minimizes its score at its optimal $\lambda$ found by the exhaustive search depicted previously on Figure~\ref{fig:lambda_selection}. This results suggests that this cross-validation method can be a viable technique for selecting the hyperparameter $\lambda$. The drawback of this approach is naturally its overhead in the sample requirement which can be far too prohibitive for some applications. An important question left to answer is to understand how these islands of consistency behave with respect to the system size and structure and if it is always judicious to choose $\lambda$ from the island of consistency corresponding to the least regularization.

\section{Scalings of the RISE with respect to $\gamma$}
\label{sec:rise_scaling}

\begin{figure}[!htb]
\begin{center}
\includegraphics[width=0.8\columnwidth]{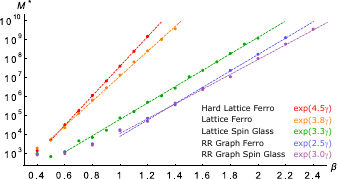}
\caption{Values of $M^{*}$ and $\gamma$-exponents for the RISE across different test cases. The hard ferromagnetic lattice is shown in red, the ferromagnetic lattice is in orange, the spin-glass on lattice is in green, the ferromagnetic random graph is in blue and the spin-glass on random graph is in purple. As for the RPLE and the logRISE, the test case requiring the larger amount of samples is the hard ferromagnetic lattice. The worst-case scaling of $M^{*}$ is $\exp(4.5 \gamma)$ which is above the information-theoretic upper bound on the optimal worse-case scaling.}
\label{fig:rise_scaling}
\end{center}
\end{figure}

The minimal amount of samples $M^*$ for a perfect reconstruction with RISE and its scaling with $\gamma$ are depicted in Figure~\ref{fig:rise_scaling}. The RISE exhibits a behavior across the different test cases that is consistent with the one shown by the RPLE and the logRISE. The hardest test case is the so called hard ferromagnetic lattice with an anti-ferromagnetic impurity introduced in the maint text for which the RISE has an exponential scaling of $4.5\gamma$, above the information-theoretical upper bound on the worse case scaling. Note also that the RISE is the best algorithm for ferromagnetic random graphs where it exhibits a scaling of $2.5\gamma$ while the RPLE and the logRISE shows $3.2\gamma$ and $3.7\gamma$ respectively. Theoretical reasons for why the RISE and the logRISE demonstrate different performance constitute an interesting research topic that remains to be studied further.

\section{High-temperature expansion of RISE and RPLE}
\label{sec:ht_expansion}

We consider the high-temperature regime for which couplings and magnetic fields are close to zero, $(J^{*}_{i},H^{*}_{i}) \sim (0,0)$, and we perform a Taylor expansion of the Pseudo-Likelihood (PL) and the Interaction Screening Objective (ISO) around this point. For PL we find the following series 
\begin{equation}
\mathcal{L}_{i}(J_{i},H_{i}) \approx - \sum_{j\neq i}J_{ij}\langle\sigma_{i}\sigma_{j}\rangle_{M} - H_{i}\langle \sigma_{i} \rangle_{M} 
+ \frac{1}{2} \sum_{j,k \neq i} J_{ij}J_{ik}\langle\sigma_{j}\sigma_{k}\rangle_{M} + H_{i}\sum_{j\neq i} J_{ij}\langle\sigma_{j}\rangle_{M} + \frac{1}{2} H^2_{i},
\label{eq:PL_taylor}
\end{equation}
and for the ISO, similarly,
\begin{equation}
S_{i}(J_{i},H_{i}) \approx 1 - \sum_{j\neq i}J_{ij}\langle\sigma_{i}\sigma_{j}\rangle_{M} - H_{i}\langle \sigma_{i} \rangle_{M} 
+ \frac{1}{2} \sum_{j,k \neq i} J_{ij}J_{ik}\langle\sigma_{j}\sigma_{k}\rangle_{M} + H_{i}\sum_{j\neq i} J_{ij}\langle\sigma_{j}\rangle_{M} + \frac{1}{2} H^2_{i}.
\label{eq:ISO_taylor}
\end{equation}
Remarkably, PL and the ISO have the exact same first and second order terms. This implies that these two estimators are equivalent in this Mean Field Regime (MFR) as their high-temperature expansion is minimized at the same value $(\widehat{J}^{\text{MFR}}_{i},\widehat{H}^{\text{MFR}}_{i})$. Taking derivatives of Eq.~\eqref{eq:PL_taylor} or Eq.~\eqref{eq:ISO_taylor} provide us with the conditions that the minimizer satifies

\begin{align}
\label{eq:taylor_first_order_a}
\widehat{J}^{\text{MFR}}_{il}\langle\sigma_{i}\sigma_{l}\rangle_{M} &= \widehat{H}^{\text{MFR}}_{i}\langle\sigma_{l}\rangle_{M}+
\sum_{j \neq i} \widehat{J}^{\text{MFR}}_{jl}\langle\sigma_{j}\sigma_{l}\rangle_{M},\\
\langle\sigma_{i}\rangle_{M} &= \widehat{H}^{\text{MFR}}_{i} + \sum_{j\neq i} \widehat{J}^{\text{MFR}}_{ij} \langle\sigma_{j}\rangle_{M}.
\label{eq:taylor_first_order_b}
\end{align}

After a little algebra, the relations Eq.~\eqref{eq:taylor_first_order_a} and Eq.~\eqref{eq:taylor_first_order_b} can be inverted to give explicit expressions of the inferred couplings and magnetic fields with respect to connected correlations and averaged magnetizations

\begin{align}
\widehat{J}_{il}^{\text{MFR}} &= -\frac{\left[\bar{C}^{-1}\right]_{il}}{\left[\bar{C}^{-1}\right]_{ii}}, \\
\widehat{H}_{i}^{\text{MFR}} &= -m_{i} + \sum_{j\neq i} \frac{\left[\bar{C}^{-1}\right]_{jl}}{\left[\bar{C}^{-1}\right]_{ii}} m_{j},
\end{align}
where the matrix of empirical connected correlations and the vector of averaged magnetizations are given by the following formulae 
\begin{equation}
\bar{C}_{ij} = \langle\sigma_{i}\sigma_{j}\rangle_M - \langle\sigma_{i}\rangle_M \langle\sigma_{j}\rangle_M, \quad m_{i} = \langle\sigma_{i}\rangle_M.
\end{equation}

\bibliography{IsingLearning1}

\end{document}